# Effect of thermal history on the short-range order in Fe-Cr alloys

S.M.Dubiel[*], J. Cieślak

AGH University of Science and Technology, Faculty of Physics and Applied Computer Science, PL-30-059 Kraków, Poland

Effect of a thermal history of $Fe_{100-x}Cr_x$ ($x < 20$) samples on a Cr atoms distribution within the first (1NN) and the second (2NN) neighbor-shells was studied with the Mössbauer spectroscopy. The distribution was expressed in terms of the Cowley-Warren short-range order (SRO) parameters: $<\alpha_1>$ for 1NN, $<\alpha_2>$ for 2NN and $<\alpha_{12}>$ for 1NN-2NN. It was shown to be characteristic of the thermal treatment and of the neigbor shell. For quenched samples, $<\alpha_1>$ is positive for all $x$-values, while $<\alpha_2>$ shows inversion at $x \approx 8$ from positive to weakly negative. Similar character has $<\alpha_{12}>$, but the degree of ordering in 1NN-2NN is lower than that in 2NN. Isochronally annealed samples exhibit similar behvior for $x > ~8$, but significantly different for $x < ~8$ where the inversion both in $<\alpha_1>$ and $<\alpha_2>$ occurs at $x \approx 3$, yet in the opposite direction. The $<\alpha_{12}>$ follows the trend predicted by Erhart et. al. [PRB 77, 134206 (2008)]. A clear-cut inversion induced by an isothermal annealing at 415 $^o$C was found for the $Fe_{85}Cr_{15}$ sample.

$Fe_{100-x}Cr_x$ alloys have been subject of intensive studies due to both their interesting physical properties [1] as well as to their industrial importance [2]. Thanks to this they are regarded and treated as model alloys for testing various models and theories. A distribution of Cr atoms has been of a particular interest, especially around a critical concentration, $x \approx 10$, above which the alloys become stainless. The existence of such critical content was found experimentally from diffuse-neutron-diffraction experiments [3,4], according to which the Cowley-Warren short-range order (SRO) parameter, $<\alpha_{12}>$, average over the first-two neighbor shells, 1NN-2NN, changes its sign from negative ($x \leq \sim 10$) – indicative of repulsion between Cr atoms – to positive ($x \geq \sim 10$) – indicative of attraction (clustering) between Cr atoms. Further experimental evidence in favor of such behavior was also found with the Mössbauer spectroscopy (MS) [5] and the synchrotron X-ray desorption technique [6]. The inversion of the SRO parameter in the Fe-Cr alloys was first predicted theoretically [7] by *ab initio* calculations. Its existence was also confirmed by calculations of the mixing entropy [8] as well as of the pair potentials applying different approaches [9-16]. It must be, however, realized that according to the atomistic Monte Carlo simulations [15], the inversion in the SRO parameter can be reproduced in thermodynamic equilibrium only then when contributions from the Fe-rich ($\alpha$) and the Cr-rich ($\alpha'$) phases are taken into account. In other words, after the phase decomposition has taken place. Otherwise, the SRO parameter for the $\alpha$ phase goes through a minimum whose position depends on the annealing temperature. These simulations clearly demonstrated that the metallurgical state of samples plays a crutial role in the actual distribution of Cr atoms, hence in the values of the SRO parameters. Indeed, our recent study on a series of Fe-Cr alloys carried out with MS gave a sound evidence of that [17]. To study the effect of samples' history in more detail, samples of $Fe_{100-x}Cr_x$ alloys, as specified in Table 1, were studied with MS, the technique that since its earliest days ($\sim$1960) has been recognized as a suitable method in studies of atomic arrangements in solids. In the case of Fe-Cr it allows to quantitatively determine the average number of Cr atoms within 1NN, $<m>$, and that within 2NN, $<m>$, hence to quantify their distribution in terms of the SRO parameters, $<\alpha_1>$ and $<\alpha_2>$, respectively, as outlined below.

**Table I.** List of investigated $Fe_{100-x}Cr_x$ samples with different thermal history: T1 stands for homogenization at 800 $^o$C, T2 for the treatment including annealing at 430 $^o$C, and T3 for annealing at 415 $^o$C. More details is given in the text. The model EFDA samples are marked with asterisk.

| No. | x | T1 | T2 | T3 |
|---|---|---|---|---|
| 1 | 2.2 | + | + | |
| 2 | 3.3 | + | + | |
| 3 | 3.9 | + | + | |
| 4 | 4.85 | + | | |
| 5* | 5.8 | + | + | |
| 6 | 6.4 | + | | |
| 7 | 7.85 | + | + | |
| 8 | 8.5 | | + | |
| 9 | 10.25 | + | | |
| 10* | 10.75 | + | + | |
| 11 | 12.3 | + | + | |
| 12 | 14.15 | + | | |
| 13 | 14.9 | + | | |
| 14* | 15.15 | + | + | + |
| 15 | 19.0 | | + | |

The samples - in form of circles (13 mm diameter) $\sim$25 $\mu$m thick – were obtained by cold rolling original ingots. Three different heat treatments were applied on the samples placed in a quartz tube:

- T1 – annealing at 800 °C for 3 h in argon, followed by quenching into liquid nitrogen.
- T2 – annealing at 800 °C for 20 h in argon, followed by 2h annealing at 520 °C. Afterwards, the temperature was slowly (20 h) decreased down to 430 °C at which temperature the samples were kept for 12 h. Finally, the quartz tube was removed quickly from the furnace and the samples thrown on a piece of brass kept in the cool zone of the tube. It should be noticed that this procedure was similar to the one used by Mirebeau and Parette [4], except the last step (samples studied in Ref. 4 were quenched into water).
- T3 – vacuum annealing at 415 °C for different periods followed by a cooling in the tube removed from the furnace.

$^{57}$Fe Mössbauer spectra were recorded at room temperature in a transmission geometry using a conventional spectrometer and a $^{57}$Co/Rh source for the 14.4 keV gamma radiation. Some examples of them recorded on the samples T1-treated are shown in Fig. 1.

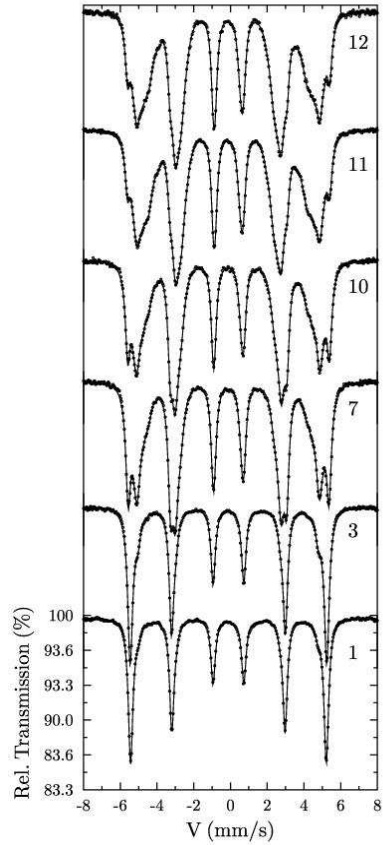

**Fig. 1** $^{57}$Fe room temperature Mössbauer spectra recorded on $Fe_{100-x}Cr_x$ samples that anderwent the treatment T1. Labels indicate the sample's number as displayed in Table 1.

The spectra were analyzed in terms of the two-shell model, assuming the effect of the presence of Cr atoms in the 1NN-2NN vicinity of the $^{57}$Fe probe nuclei on the hyperfine field (*B*) and on the isomer shift (*IS*) is additive i.e. $X(m,n) = X(0,0) + m \cdot \Delta X_1 + n \cdot \Delta X_2$, where *X = B* or *IS*, and $\Delta X_i$ is a change of *B* or *IS* due to one Cr atom situated in 1NN (*i=1*) or on 2NN (*i=2*). Twenty five most significant atomic configurations, *(m,n)*, taken into account were chosen based on the binomial distribution. However, their probabilities, *P(m,n)*, were treated as free parameters (their starting values were those from the binomial distribution). All the spectra were fitted simulateneously with a least-squares method assuming the same values of $\Delta X_i$'s. All other spectral parameters like *X(0,0)*, linewidths of individual sextets *G1*, *G2* and *G3* and their relative intensities (Clebsch-Gordan coefficients) *C2* and *C3* were treated as free ($C1$=1). Very good fits (in terms of $\chi^2$) were obtained with the following values of the spectral parameters: $\Delta B_1$= -3.05 T, $\Delta B_1$= -1.95 T, $\Delta IS_1$= -0.020 mm/s, $\Delta IS_1$= -0.007 mm/s, $G1$=0.28(2) mm/s, $G2$=0.30(2) mm/s, $G3$=0.32(2) mm/s, $C2$=2.2(4), $C3$=2.5(1). The knowledge of the atomic configurations, $(m,n)$, and their probabilities, $P(m,n)$, permited to determine $<m> = \sum_{m,n} m P(m,n)$, $<n> = \sum_{m,n} n P(m,n)$, and $<m+n> = \sum_{m,n} (m+n) P(m,n)$.

Knowing $<m>$, $<n>$, and $<m+n>$, was enough for calculation of the corresponding SRO parameters $<\alpha_1>$, $<\alpha_2>$, and $<\alpha_{12}>$. For that purpose the following equations were used:

$$\langle \alpha_1 \rangle = 1 - \frac{\langle m \rangle}{\langle m_r \rangle} \quad (1a)$$

$$\langle \alpha_2 \rangle = 1 - \frac{\langle n \rangle}{\langle n_r \rangle} \quad (1b)$$

$$\langle \alpha_{12} \rangle = 1 - \frac{\langle m+n \rangle}{\langle m_r + n_r \rangle} \quad (1c)$$

Where $<m_r>$=0.08x, $<n_r>$=0.06x, and $<m_r+n_r>$=0.14x are the average numbers of Cr atoms in 1NN, NN, and 1NN-2NN, respectively, as expected for the random distribution. The $<\alpha_i>$-values obtained in that way are plotted in Fig. 2.

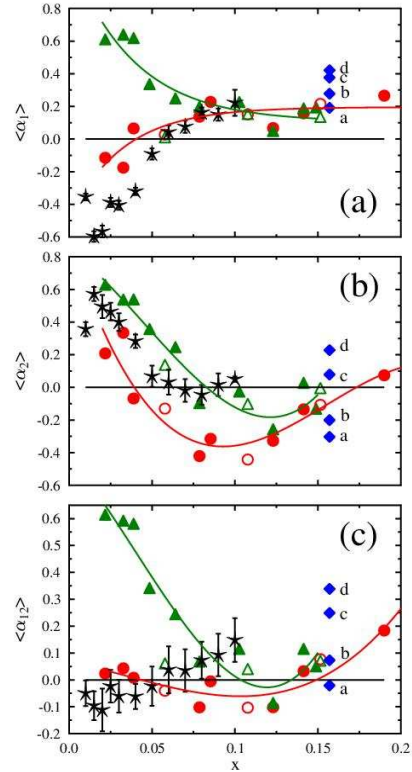

**Fig. 2** SRO parameters versus Cr content, $x$, as calculated for (a) 1NN-shell, (b) 2NN-shell and (c) 1NN-2NN- shells. Data from Ref. 18 (asterisks) are added for comparison. Solid lines are marked to quide the eye. Trianles represent the data obtained for the treatment T1, circles are for T2 and diamonds for T3 with different annealing times: 0h (a), 10h (b), 192h (c) and 840 h (d) . Open symbols depict the model EFDA samples.

It is clear that the SRO parameters depend both on the thermal history and composition in a way characteristic of the neighbor shell. Concerning the 1NN-shell, all $<\alpha_1>$-values are positive for T1 indicating thereby a clustering of Cr atoms. The degree of the clustering being higher for smaller $x$ what is rather unexpected as high-temperature annealing follwed by quenching is believed to result in a random distribution of atoms. For the samples that underwent T2 the behavior of $<\alpha_1>$ is more complex as it shows an inversion at $x \approx 3$ from weak negative to weak positive, and, for higher concentration. it hardly depends on $x$. The sample treated with T3 has positive $<\alpha_1>$ that amplitude, hence the degree of clustering, gently increases with the annealing time. The behavior of the data taken from the literature [18] are similar to that found for T1, except the inversion takes place at $x \approx 6$ and the degree of ordering below $x \approx 6$ is higher (these samples were annealed at 997 °C for 2h and then during 6h cooled down to room temperature). Regarding the 2NN-shell, $<\alpha_2>$ shows inversion from positive to negative for both T1 and T2 viz. at $x \approx 3$ for T2, and at $x \approx 8$ for T1. In the range of negative $<\alpha_2>$-values both for T1 and T2 there are minima characteristic of the treatment, and for $x>\sim12$ the behavior hardly depends on the treatment, and for the both cases it shows a tendency towards a second inversion at $x \approx 15$. For the sample treated with T3 $<\alpha_2>$ is initially negative, but its amplitude decreases with the annealing time and eventually becomes positive. In other words, one observes here a phase-decomposition induced inversion. The data from Ref. 18 also exhibits some anomaly at $x \approx 6$. At higher contents $<\alpha_2> \approx 0$. Finally, concerning the SRO parameter averaged over both shells, $<\alpha_{12}>$, it shows for T1 a steap decrease with $x$ from large positive values, hence from a high degree of clustering, to about zero at $x \approx 12$. The latter means that the departure from randomness within the 1NN-2NN volume for $x>\sim12$ is moderate, if any. On the other hand, the behavior of $<\alpha_{12}>$ for T2 resembles the one predicted by the Monte Carlo simulations i.e. $<\alpha_{12}>$-values are weakly negative with a shallow minimum [15]. An inversion can be observed at $x \approx 15$. The average SRO parameter for T3 was initially close to zero, hence the distribution of Cr atoms over 1NN-2NN was random, but on annealing it became clearly positive, hence indicative of the clustering. Finally, the literature data shows a weak inversion at $x \approx 5$ indicating thereby a transition from the ordering to clustering of Cr atoms.

The data obtained in this study and that reported in the literature [3-5,17,18] give a clear-cut evidence that the actual distribution of Cr atoms in Fe matrix meaningfully depends on thermal history of the samples. In particular, the concentration at which the inversion of the SRO parameter occurs shows such dependence. A rather unexpected finding is that the most significant deviation from randomness was revealed in low-concentrated alloys that underwent a homogenisation treatment. As evidence in Fig. 2, all three SRO parameters are positive for $x<\sim7$ i.e. the custering of Cr atoms occurs. In other words, in that range of composition, the average number of Cr atoms in the 1NN-2NN vicinity of the probe $^{57}$Fe nuclei, $<m+n>$, is reduced relative to that expected from the binomial distribution. The reduction of $<m+n>$ is equivalent to an increase of $<B>$ [19], hence the behavior of $<B>$ can be taken as a proper measure for the SRO-related effects. Here, for $x<\sim7$, $<B>$ has, for the T1-treated samples, higher values than the corresponding ones expected for the random distribution – see Fig.3. This can be regarded as an independent evidence in favour of the clustering of Cr atoms in that composition range.

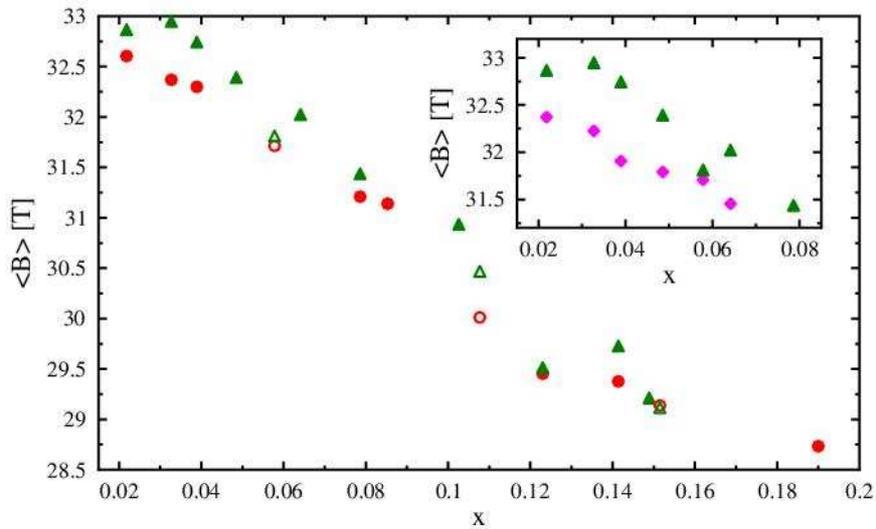

Fig. 3 The average hyperfine field, <B>, versus Cr contents, x, as determined for the samples treated with T1 (triangles) and T2 (circles). Diamonds stay for <B> expected for the random distribution.

In summary, an experimental evidence was found that the SRO parameters are characteristic of a given atomic shell (1NN, 2NN), and for a given shell they significantly depend on the applied heat treatment. The difference is especially large for $x<\sim7$. The homogenized samples, contrary to the expectation, do not have random distribution of Cr atoms, the departure from randomness being particularly big for low-concentrated samples where a high degree of clustering was revealed. The SRO $<\alpha_{12}>$ - parameter of the samples in the thermodynamic equilibrium (heat treatment T2) resembles that predicted with the Monte Carlo simulations [15]. An evidence of a clear-cut inversion was found for the $Fe_{85}Cr_{15}$ sample that underwent a prolonged isothermal annealing at 415 $^o$C.

**Acknowledgements**

The study was carried out within the 7[th] Frame EU Programme EURATOM and was supported by the EFDA-IPPLM Association and Polish Ministry for Science and Higher Education, Warszawa.

* Corresponding author: Stanislaw.Dubiel@fis.agh.edu.pl